\documentclass[11pt]{article}
\pdfoutput=1
 \usepackage{mciteplus}
 \usepackage{tikz}
 \usepackage{color}
 \usepackage{xcolor}
 \usepackage{comment}
 \definecolor{darkblue}{rgb}{0.1,0.1,.7}
 \usepackage[colorlinks, linkcolor=darkblue, citecolor=darkblue, urlcolor=darkblue, linktocpage,hyperfootnotes=false]{hyperref} 
\usepackage{epsfig}
\usepackage{graphicx}
\usepackage{cite}
\usepackage{amsfonts}
\usepackage{amssymb}
\usepackage{bm}
\usepackage{latexsym}
\usepackage{mathtools}
\usepackage{amsmath}
\numberwithin{equation}{section}
\def\bq{\begin{quote}}
\def\eq{\end{quote}}
 at 10truept

\newcommand{\cala}{{\cal A}}
\newcommand{\calc}{{\cal C}}
\newcommand{\calm}{{\cal M}}
\newcommand{\caln}{{\cal N}}
\newcommand{\calo}{{\cal O}}
\newcommand{\calh}{{\cal H}}

\newcommand{\calx}{{\cal X}}

\newcommand{\beq}{\begin{equation}}
\newcommand{\eeq}{\end{equation}}
\newcommand{\beqa}{\begin{eqnarray}}
\newcommand{\eeqa}{\end{eqnarray}}
\newcommand{\bea}{\begin{eqnarray}}
\newcommand{\eea}{\end{eqnarray}}

\newcommand{\hf}{\frac{1}{2}}

\def\roughly#1{\raise.3ex\hbox{$#1$\kern-.75em\lower1ex\hbox{$\sim$}}}

\begin{document}

\thispagestyle{empty}
\begin{titlepage}
  \bigskip

  \bigskip\bigskip

  \bigskip

\begin{center}
{\Large \bf {Challenges for describing unitary evolution in nontrivial geometries: pictures and representations
}}
    \bigskip
\bigskip
\end{center}

  \begin{center}

 \rm {Steven B. Giddings\footnote{\texttt{giddings@ucsb.edu}} and Julie Perkins\footnote{\texttt{jperkinsphd@gmail.com}}}
  \bigskip \rm
\bigskip

{Department of Physics, University of California, Santa Barbara, CA 93106, USA}  \\
\rm

  \bigskip \rm
\bigskip
 
\rm

\bigskip
\bigskip

  \end{center}

\vspace{3cm}
  \begin{abstract}
 \medskip
  \noindent
  Description of evolution between spatial slices in a general spacetime suffers from a significant difficulty: the states on the slices, in a given basis, are not related by a unitary transformation.  This problem, which occurs in spacetime dimensions above two, is directly related to the infinite number of inequivalent representations of the canonical commutators, and in particular will arise for interacting theories in time-dependent spacetimes.  We connect different facets of this issue, and discuss its possible resolution.  It is directly related to discussions of failure of a standard Schr\"odinger picture of evolution, and of evolution via ``many-fingered time." One requires a condition specifying a physical unitary equivalence class of states; in general this equivalence class evolves with time, and an important question is how it is determined.  One approach to this in free theories is by imposing a Hadamard condition on the two point function.  We explore a different approach, which also may be helpful for interacting theories, analyzing the structure of the state in a local limit, and relate these approaches.  We also elucidate the non-Hadamard behavior of unphysical vacua, and  discuss concrete examples of these approaches involving cosmological and black hole evolution.  The issues are extended in the context of quantum dynamical geometry, and raise important questions for the proper description of the wavefunction of the universe and for the role of  the Wheeler-DeWitt equation.
  \end{abstract}
\bigskip \bigskip \bigskip 

  \end{titlepage}

\section{Introduction and motivation}

For quantum field theory (QFT) in flat space, and for other quantum mechanical systems, it is very useful to be able to describe evolution in different pictures, {\it e.g.} Heisenberg and Schr\"odinger, or intermediate interaction pictures.  For example if the focus is on evolution of observables, Heisenberg picture, where operators are time dependent but the quantum state is fixed, gives the most transparent description.  Or, if one's focus is on evolution of the quantum state, Schr\"odinger picture captures that with an evolving quantum state and fixed operators.  In particular, with the more modern focus on aspects of quantum information -- its localization, its flow, and evolving entanglement, a description of the evolving quantum state seems to be the approach that makes these features most transparent.  

Including gravity -- beginning with introducing nontrivial background spacetimes, and going beyond with dynamical geometry -- adds considerable subtlety to this story.  Specifically, there are known issues\cite{Helf,ToVa1,ToVa2,CMOV,CFMM,AgAs,KoMo} in giving a traditional Schr\"odinger picture description in a general curved background geometry, and these appear to be magnified when that geometry is dynamical.  Despite these, there is a widely adopted view that one can simply describe unitary evolution of the quantum state from slice to slice in a background geometry.  However, as will be discussed below, the description of such evolution encounters significant subtleties.  One reason this may come as a surprise, particularly to the large segment of the community who have developed their intuition for curved geometries from the worldsheets of string theory, is that this difficulty arises for spacetime dimension $D>2$, and is not a problem for two-dimensional geometries.

Of course different facets of this problem are known to those who have carefully studied higher-dimensional QFT in curved backgrounds.  This and related problems are part of the motivation for adopting an algebraic approach, generalizing a Heisenberg picture description, as opposed to using traditional canonical methods.  However, there are drawbacks to being restricted to such an approach.  First, connections are most transparently made to discussions of quantum information theory if one works with a description of the evolving quantum state, rather than with a fixed state and evolving operators.  Second, for an interacting theory, the evolution of the operators is governed by the generalized Heisenberg equations, which are then nonlinear and very difficult to solve, or even precisely formulate; this suggests a possible utility for describing instead the evolution of the state.  Finally, this complex of issues seems to raise even more difficult issues with dynamical quantum geometry; for example, there is a sense in which the Wheeler-DeWitt equation is more closely tied to a Schr\"odinger picture description.

In fact, the underlying problem appears to extend beyond that of continuous evolution to a problem for describing even the basic transition amplitudes between initial and final quantum states in a general geometry, which we might have expected to be basic quantities in a quantum theory.  And, in general spacetimes, even a diffeomorphism produces a nonunitary map between Hilbert spaces\cite{ToVa2}.

One goal of this paper is to investigate these questions, and work towards a simplified account and deeper understanding of the connection between the abstract algebraic (``Heisenberg") approach and a more traditional canonical (``Schr\"odinger") approach, and of the relevant obstacles.  

In approaching the problem from the canonical side, it is found to be directly connected to the known problem of there being infinitely many unitarily {\it inequivalent} representations of the basic commutators of the Heisenberg algebra for fields.  In brief, if we begin with an initial state in a given representation, and consider its evolution,
the resulting state is not necessarily unitarily equivalent to one in a representation that we might choose to define on a subsequent slice.  This raises the question of how a ``good," or physical, representation for the quantum state is fixed.

In outline, we begin by describing canonical quantization of a free field on a background geometry.  In order to construct the quantum states, {\it e.g.} in a Fock basis, certain choices must be made, which amount to the choice of a complex structure on the phase space.  We describe the different choices, and the possibility that in QFT these are not necessarily unitarily equivalent.  This leads to the notion of {\it equivalence classes} of such complex structures that {\it are} unitarily equivalent.  We then elaborate on a basic example\cite{ToVa2} that shows that evolution in general takes one out of a given equivalence class, and which gives an example of the problem with Schr\"odinger-like evolution, explicitly showing the role of $D>2$.

A question then becomes, what chooses a physical representation, or, equivalence class of representations?  A previous characterization of this in an algebraic approach has been in terms of the singular behavior, specifically of the Hadamard form, of the correlation functions\cite{Hada,Waldbook}.  We instead begin section three with a local analysis of the description of the quantum state on a given slice, seeking a complementary approach avoiding issues noted above with the algebraic approach.  Specifically, studying the problem in the ADM\cite{ADM} framework and introducing a local Lorentz frame and description, matching onto the locally flat structure of the spacetime manifold, leads to criteria defining physical representations.  This specifically depends on the background geometry and slicing through the embedding metric on the slices.
The resulting conditions can be related to the Hadamard criterion; in simple examples one finds that physical representations yield Hadamard behavior of the two-point functions, and conversely, unphysical representations produce singular behavior not of Hadamard form.  This also gives a possibly useful perspective on the general problem (see {\it e.g.} \cite{Kayrev}) of constructing Hadamard states in general spacetimes.

Section four then describes examples of this discussion, tying together some known examples in the literature.  Specifically, we consider simple
cosmologies, and evolution in black hole backgrounds.  In the latter, we can connect to a description of unitary evolution on stationary slices\cite{NVU,SEHS,GiPe1}

Section five concludes with a more general discussion of the basic situation, and further problems and questions.  This begins with overviewing the case of free theories on fixed backgrounds, which is the main subject of this paper.  Interacting theories produce additional difficulties, as noted above, which it may be advantageous to study in the approach of this paper via a canonical approach and local analysis.  Finally, we discuss the problem of dynamical geometries, and new issues that arise there in attempting to define the wavefunction of the Universe and the constraints ({\it e.g.} Wheeler-DeWitt equation) that fix it.  This presents challenges to the description that has been called ``many-fingered time."  Here, also, one encounters the representation problem, but with an added difficulty that the representations depend on the dynamical geometry of the metric, or whatever replaces it in the full quantum theory, leaving the bigger question of how to proceed with dynamical quantum geometry.  

An appendix describes quantization in finite volume, or in local finite volumes, which is utilized in the main text.

\section{Quantization, representations, and equivalence classes}

Our focus will be on the simple case of a real scalar field; we will defer the additional complications of fields with spin for future work.  On a $D$-dimensional manifold with background metric, $(\calm, g_{\mu\nu})$, the action is
\beq\label{scalact}
S=-\hf \int d^D x \sqrt{|g|}\left[ (\nabla\varphi)^2 +m^2\varphi^2\right]\ .
\eeq
We begin with a description of its canonical quantization, and associated subtleties.

\subsection{Canonical quantization and complex structures}
\label{CanonQ}

To canonically quantize, we introduce a foliation of $\calm$ by spacelike slices, which we parameterize as $x^\mu={\cal X}^\mu(t,y^i)$.  Here $t$ is a time parameter labeling slices of the foliation, and for a given $t$, ${\cal X}^\mu$ can be thought of as mapping a spatial manifold $\Sigma$, with coordinates $y^i$, into $\calm$.  The canonical momentum (defined as a density) is 
\beq\label{pidef}
\Pi= \frac{\delta S}{ \delta \dot\varphi} ={\sqrt q} \partial_n\varphi\ ,
\eeq
with $\sqrt q$ the metric induced by $g$ on $\Sigma$ and $\partial_n\varphi=n^\mu\partial_\mu \varphi$ the derivative normal to the embedded slice.  Then, the canonical commutation relations are
\beq\label{CCRs}
[\Pi(t, y^i), \varphi(t,y^{i\prime})]=-i\delta^{d}(y-y')\ ,
\eeq
together with vanishing equal time $[\varphi,\varphi]$ and $[\Pi,\Pi]$, and with $d=D-1$.
These are assumed to be extendable to unequal times by solving the quantum equations of motion for the operators $\varphi$ and $\Pi$.  

This allows us to define an algebra $\cala$ of quantum operators generated by $\varphi$ and $\Pi$.\footnote{In a more rigorous treatment, one typically exponentiates integrals of $\varphi$ and $\Pi$ to give bounded operators.}  In order to define the quantum theory, one also needs to define a {\it representation} of the algebra arising from \eqref{CCRs}, acting on quantum states.  This leads to additional subtlety, since it turns out that there are an infinite number of inequivalent representations of the algebra\cite{GaWi},\cite{Waldbook,WittQFT}.

In order to study these representations, first note that the classical data $\gamma(y)=(\phi(y), \pi(y))$ specifies a point in the phase space $\Gamma_{\rm can}$, and for an embedding ${\cal X}^\mu(t,y^i)$ of $\Sigma$ that is a Cauchy slice, specifies initial data at $t$ identifying this point with a solution $\varphi(x)$ to the classical field equations on $\calm$, via $\varphi(t,y^i)=\phi(y^i)$, $\Pi(t,y^i)=\pi(y^i)$.\footnote{To avoid notational clutter, we follow the common practice of using the same symbols for the classical fields, and for the corresponding quantum operators; a careful reading of context should help avoid confusion.}  There is a symplectic product between such solutions
\beq\label{prodsoln}
\Omega(\varphi_1,\varphi_2) = \int_S d^{d}x\sqrt{q_S}n_S^\mu(\varphi_1  \partial_\mu\varphi_2 -  \partial_\mu\varphi_1  \varphi_2)\ ,
\eeq
where $S$ denotes a Cauchy slice in $\calm$, with corresponding induced metric and normal, which could {\it e.g.} be an image of $\Sigma$; this product is conserved by the equations of motion.  This induces the symplectic structure
\beq\label{symps}
\Omega(\gamma_1,\gamma_2) =\int _\Sigma d^{d}y (\phi_1 \pi_2 -\pi_1 \phi_2) 
\eeq
on $\Gamma_{\rm can}$.  

One can then proceed towards a Fock space construction of the states by introducing a basis $\gamma_I=(\phi_I,0)$, $\gamma_{I'}=(0,\pi_{I'})$ for $\Gamma_{\rm can}$.  In order to avoid infrared issues associated with infinite volume,\footnote{For further recent discussion of these issues, see \cite{WittQFT}. Finite volume quantization conventions are summarized in Appendix~\ref{appa}, and for the case where $\Sigma = {\mathbb R}^{d}$, we may formally take the infinite volume limit using $\sum_I\rightarrow \frac{V}{(2\pi)^{d}}\int d^{d} k$, {\it etc}. } we will treat the case of compact $\Sigma$, and so take $I,I'$ to be discrete indices.  These will have infinite range, but can be restricted to finite range, $I,I' = 1,\ldots,N$, by an appropriate UV cutoff.  In order to simplify, parts of our discussion will be given with such a finite $N$, which we may then take to infinity by removing the cutoff.

The definition \eqref{symps} implies that, evaluating on the basis, $\Omega_{IJ}=\Omega_{I'J'}=0$; 
by a Graham-Schmidt orthogonalization procedure, one may pass to a {\it symplectic basis} in which
\beq\label{sympform}
\Omega_{II'}=\int d^{d}y \phi_I \pi_{I'} = \delta_{II'} = -\Omega_{I'I}\ .
\eeq
Two such symplectic bases are related by an $Sp_{2N}(\mathbb R)$ transformation, leaving this form invariant.\footnote{More extensive discussion of symplectic geometry may be found in \cite{Berndt}.}

A Fock space representation of the quantum states is constructed by defining a notion of ``positive frequency" solutions; these are determined\cite{AsMa1,AsMa2,AgAs} by specifying a complex structure $J$ that acts as an automorphism on $\Gamma_{\rm can}$, satisfies $J^2=-1$, and is taken to be {\it compatible} with the symplectic structure; $\Omega(J\gamma_1,J\gamma_2)= \Omega(\gamma_1,\gamma_2)$.  An example  is the ``standard" $J$, given by
\beq\label{stanJ}
J_0=\begin{pmatrix}
0&-I\\
I&0
\end{pmatrix}
\eeq
in the symplectic basis.  Then, one may define projectors
\beq
P_{\pm}= \frac{1\mp i J}{2}
\eeq
which project a given $\gamma(y)\in\Gamma_{\rm can}$ (or corresponding solution) onto what we call its positive or negative frequency part.  The resulting positive and negative frequency subspaces are $N$-dimensional, and may be taken to have  complex bases $\gamma_A$, $\gamma_A^*$, respectively, with $A=1,\ldots,N$.   A symmetric form may be defined on $\Gamma_{\rm can}$ by
\beq
\langle \gamma_1,\gamma_2\rangle = \Omega(\gamma_1, J\gamma_2)\ ;
\eeq
symmetry follows from the compatibility of $J$ with $\Omega$.  For $J$ such that $\langle\gamma,\gamma\rangle$ is positive definite, this defines an hermitian metric, and $(\Gamma_{\rm can}, \Omega, J)$ together define a K\"ahler space.
The compatibility condition also  implies $\langle \gamma_A, \gamma_B\rangle =0$; the nonzero elements
\beq
\langle\gamma_A,\gamma^*_B\rangle = G_{A\bar B}
\eeq
may be set to $G_{A\bar B}=\delta_{AB}$ by a Graham-Schmidt procedure.

With these structures in place, we now consider quantum operators $\gamma(y)=(\phi(y),\pi(y))$, defined on $\Sigma$ and satisfying the commutation relations corresponding to  \eqref{CCRs}
\beq\label{CCRy}
[\phi(y),\pi(y')] = i \delta^{d}(y-y')\ .
\eeq
We then define the annihilation and creation operators in terms of the expansion
\beq\label{ladderexp}
\gamma(y) = \sum_A (a_A \gamma_A + a^\dagger_A \gamma_A^*)\ ,
\eeq
where for completeness we now need to take $N\rightarrow\infty$.
Alternately, from the preceding definition of the inner product,
\beq\label{modprod}
a_A=\langle \gamma_A^*, \gamma\rangle\ ,\ a^\dagger_A=\langle \gamma_A, \gamma\rangle\ .
\eeq
Then the commutation relations \eqref{CCRy} and vanishing of $[\phi,\phi]$ and $[\pi,\pi]$ imply 
\beq
[a_A,a^\dagger_B]=\delta_{AB}\ ,
\eeq
together with vanishing $[a_A,a_B]$ and $[a_A^\dagger,a_B^\dagger]$.
The construction of the Fock space of states then immediately follows; the vacuum state $|0\rangle$ is defined to be the state annihilated by all the $a_A$, and other states are found by acting with superpositions of products of the $a_A^\dagger$'s on the vacuum.  This furnishes an explicit representation of the algebra $\cala$.

\subsection{Representations and equivalence classes}

The preceding definitions involved certain choices.  Suppose we had chosen a different symplectic basis, related to the original one by an $Sp_{2N}(\mathbb R)$ transformation $O$.  Then, if we had chosen the standard complex structure $J_0$ in the new basis, that would correspond to the complex structure $J=O^{-1}J_0 O$ in the original basis, and would  in general yield a new complex structure.

However, not {\it all} such $O\in Sp_{2N}(\mathbb R)$ give us a new complex structure; equivalent complex structures are related by transformations $\gamma_A\rightarrow U_{AB}\gamma_B$, with $U\in U(N)$, which mix positive frequency modes.  As a result, one can show (see {\it e.g.} \cite{Berndt}) that the space of compatible complex structures (also satisfying the positivity condition) can be identified with $Sp_{2N}(\mathbb R)/U(N)$, also known as the Siegel upper half plane.  And, different complex structures then correspond to different definitions of the basic annihilation and creation operators, and hence of the vacuum, and thus different representations of $\cal A$.  

The case $N=1$ furnishes an illustrative example.  The symplectic form \eqref{sympform} with $N=1$ is left invariant by
\beq
\Omega \rightarrow O\Omega O^T
\eeq
where
\beq
O=
\begin{pmatrix}
a&b\\
c&d
\end{pmatrix}
\eeq
with $ad-bc=1$, also exhibiting the equivalence $Sp_{2}(\mathbb R) \simeq SL(2,\mathbb R)$.  Starting with commutators $[\phi,\pi]=i$, the standard $J_0$ of \eqref{stanJ} gives 
\beq \label{adef}
a= \frac{1}{\sqrt 2} (\phi+i\pi)\ , a^\dagger= \frac{1}{\sqrt 2} (\phi-i\pi)\ .
\eeq
If we choose a new basis defined by acting with $O$ on the original basis, and use the standard complex structure in that basis, then in the original basis that will correspond to a new complex structure $J=O^{-1} J_0 O$ as described above.  A short calculation then shows that the corresponding ladder operators associated to the new $J$ are of the form
\beq\label{newa}
\tilde a = \frac{e^{i\alpha}}{\sqrt{2Im \tau}} (\phi + \tau \pi)\ ,\  \tilde a^\dagger = \frac{e^{-i\alpha}}{\sqrt{2Im \tau}} (\phi + \bar \tau \pi)\ ,
\eeq
where
\beq \label{atdef}
\tau = \frac{b+id	}{a+ic}\ .
\eeq
Here $\alpha$ is a phase that is irrelevant to defining the complex structure, and can be dropped, as a special case of the above quotient by $U(N)$.  Combining \eqref{adef}  and \eqref{newa}  gives
\beq \label{1dbog}
\tilde a =  \frac{e^{i\alpha}}{2\sqrt{Im \tau}}\left[(1-i\tau)a + (1+i\tau) a^\dagger\right]\ .
\eeq
Thus the different ladder operators corresponding to the different complex structures are related by a Bogolubov transformation, and in particular the vacuum $|\tilde 0\rangle$ defined by the new complex structure is found by acting with this Bogolubov transformation on the original vacuum $|0\rangle$.  This transformation provides a unitary map between the representations corresponding to the two different complex structures. Since the original choice of basis was arbitrary, the transformation \eqref{1dbog} corresponds to the general such relation between complex structures.

The $N>1$ case generalizes this discussion.  A change in the complex structure,  {\it e.g.} induced by a change in symplectic basis for $\Gamma_{\rm can}$, leads to a Bogolubov transformation
\beq\label{Bog}
 a_{A} = A_{AB} \tilde a_B + B^*_{AB} \tilde a_B^\dagger\ ,\ a_{A}^\dagger = A_{AB}^* \tilde a_B^\dagger + B_{AB} \tilde a_B\ ,
\eeq
with
\beq
A A^\dagger - B^* B^T =1\quad , \quad AB^\dagger-B^*A^T=0\ .
\eeq
For finite $N$, the resulting transformation can be implemented by constructing the unitary operator relating  the $\tilde a_{A},\tilde a_{A}^\dagger$'s to the $a_{A}, a_{A}^\dagger$'s; for example, this in general relates $|\tilde 0\rangle$ to a superposition of excited states in the original Fock space based on $|0\rangle$, given by\cite{Stone,PPR}
\beq\label{Bog0}
|\tilde 0\rangle = \caln \exp\left\{\hf a_A^\dagger X_{AB} a_B^\dagger \right\} |0\rangle\ ,
\eeq
with 
\beq
X= (A^\dagger)^{-1} B^\dagger\quad ,\quad \caln=det^{-1/4}(AA^\dagger)=det^{-1/4}\left(1+B^*B^T\right) .
\eeq

However, a new issue arises in the limit of infinite $N$.  This can for example be seen by considering the expectation value of the original number operator in the new vacuum,
\beq\label{Bsquared}
\langle \tilde 0| \sum_A a_A^\dagger a_A |\tilde 0\rangle = \sum_{AB} |B_{AB}|^2\ .
\eeq
This expression may diverge, signaling the absence of a unitary map relating the Fock spaces based on the two different complex structures.  The problem may be alternately seen\cite{Wald1979} by considering, for example, the norm of the two-particle component of $|\tilde 0\rangle$, or the overlap $\langle 0|\tilde 0\rangle$, that both follow from \eqref{Bog0}, and whose finiteness depends on that of \eqref{Bsquared}.  If we define 
\beq
|\tilde\psi_2\rangle = \hf a_A^\dagger X_{AB} a_B^\dagger |0\rangle\ 
\eeq
then $\langle\tilde\psi_2|\tilde\psi_2\rangle$ diverges when \eqref{Bsquared} diverges, and correspondingly the overlap $\langle 0|\tilde 0\rangle$ vanishes.

Such divergences and the absence of a unitary map correspond to the failure of the operator $B$ to be Hilbert-Schmidt\cite{Shale}. When this happens, 
the $a$ and $\tilde{a}$ bases each correspond to one of infinitely many unitarily inequivalent quantum field theory constructions in a curved spacetime. 
This behavior is the origin of the  infinite number of unitarily inequivalent representations of the basic commutators \eqref{CCRs}. 
Without a preferred basis, the problem is not the validity of each state, but the lack of a suitable transformation between them.
For time-dependent spacetimes there is, in general, no natural definition of  preferred state, as different slicings of it correspond to different sets of positive and negative frequency modes.  As we will explore further, a consequence of this is that the  unitary evolution operator $U(t)$ of the Schr\"odinger picture cannot be defined in the usual way.

In order to characterize this behavior, we define an {\it equivalence} of two complex structures, $J_1\simeq_F J_2$, when there {\it is} such a unitary (Hilbert-Schmidt) map between  their corresponding Fock spaces.  This then defines a notion of {\it equivalence classes} $\{ J\}$ of complex structures, within which there are such unitary maps. 

\subsection{Evolution of equivalence classes}\label{EvolEC}

To recapitulate the preceding discussion, given a complex structure on the one particle phase space $\Gamma_{can}$, one may give a Fock space construction of a Hilbert space of states, and given an embedding   $\calx^\mu(t,y)$ of $\Sigma$ into $\cal M$ at fixed $t$, these may be thought of as quantum states on the corresponding slice of $\cal M$.  Suppose that one then considers the evolution on $\cal M$ to another slice, given by $\calx^\mu(t',y)$. Of course, we expect the resulting state to correspond to a different complex structure; this is the statement that in general there is particle production, so for example the initial state $|0\rangle$ evolves to an excited state on a later slice.  But is the resulting state in the same {\it equivalence class} of complex structures, that is, can it be unitarily related to the initial state?  This is what one would expect for Schr\"odinger evolution in the traditional form.
The answer, as demonstrated in \cite{Helf,ToVa1, ToVa2}, is for $D>2$ {\it no}: in general evolution cannot be thought of as taking place within the ``same" Hilbert space, in this sense, and instead leads to evolution between different equivalence classes.  Put differently, the representations of the commutators \eqref{CCRy} at $t$ and $t'$ are in general inequivalent.

This is demonstrated by \cite{ToVa2} in the special case where $\cal M$ is Minkowski space, or Minkowski space with a periodic identification of spatial directions with common length $L$ to make $\Sigma$ compact.  A particularly simple case takes a flat initial slice, $x^\mu=\calx^\mu(t_i,y^i)=(0,y^i)$, but a nontrivial final slice $x^\mu=\calx^\mu(t_f,y^i)$, where $x^\mu$ are the Minkowski coordinates.  In that case, one can define a symplectic basis and the standard canonical structure at $t_i$ to be associated with the standard positive frequency solutions\footnote{For finite volume normalization conventions, see Appendix~\ref{appa}.}
\beq\label{MPF}
\varphi_k = \frac{1}{\sqrt {2\omega_k V}} e^{ik_\mu x^\mu}\ 
\eeq
with $V=L^d$, $k_\mu=(-\omega_k, k_i)$, 
\beq\label{omegak}
\omega_k=\sqrt{k_i^2+m^2}\ ,
\eeq
and so correspondingly
\beq
\label{flatdata}
\gamma_k=(\phi_k,\pi_k)= \frac{1}{\sqrt{2\omega_kV}} \left(e^{i k_i y^i}, -i\omega_k e^{i k_i y^i}\right)\ .
\eeq
The positive frequency solutions \eqref{MPF} evolve to Cauchy data
\beq
\label{slicedat}
\tilde\gamma_k=\frac{1}{\sqrt{2\omega_kV}}\left(e^{ik_\mu\calx^\mu(t_f,y)}, i\sqrt q_f n^\mu_f k_\mu e^{ik_\mu\calx^\mu(t_f,y)} \right)
\eeq
at $t_f$, where $q_f$ and $n_f$ denote the corresponding induced metric and normal, and so the {\it evolved} complex structure is associated with the expansion
\beq
\gamma = \sum_k( \tilde a_k \tilde \gamma_k +  \tilde a_k ^\dagger \tilde \gamma_k^*)\ .
\eeq
In contrast, the original complex structure is associated with the expansion
\beq
\gamma = \sum_k( a_k  \gamma_k + a_k ^\dagger \gamma_k^*)\ .
\eeq

From \eqref{modprod}, the Bogolubov coefficients \eqref{Bog} are given by
\beq\label{ABdef}
A_{kk'}=\langle\gamma^*_k,\tilde \gamma_{k'}\rangle\ ,\ B_{kk'}= \langle\gamma_k,\tilde \gamma_{k'}\rangle\ ,
\eeq
and explicitly,
\beq\label{betaf}
B_{kk'}=- \frac{1}{\sqrt{2\omega_k 2\omega_{k'}}V} \int d^{d}y \left( \omega_k+ \sqrt{q_f} n^\mu_f k'_\mu\right) e^{i k_i y^i +i k_\mu' \calx^\mu(t_f,y^i)}\ .
\eeq
The ultraviolet behavior of \eqref{betaf} may be estimated\cite{ToVa2} via the stationary phase approximation.  Consider large $k$ with fixed $k'/k$; here the exponent in  \eqref{betaf} will be rapidly oscillating.  As a result we expect the integral to be dominated by one or more stationary points of the phase 
\beq
\Phi=k_i y^i + k_\mu' \calx^\mu(t_f,y^i)\ , 
\eeq
given by $\partial \Phi/\partial y^i =0$.  The leading contribution at such a point will have an inverse factor of $det(\partial_i\partial_j\Phi)^{1/2}$, producing a power $1/k^{d/2}$ in the integral.  For nonvanishing coefficient $C=(\omega_k+ \sqrt{q_f} n^\mu_f k'_\mu)/\sqrt{2\omega_k 2\omega_{k'}}$, we thus find behavior $B_{kk'}\sim 1/k^{d/2}$ in this limit.  The UV contribution to the sum \eqref{Bsquared} then behaves as
\beq\label{divsum}
 \sum_{kk'} |B_{kk'}|^2 \sim V^2 \int d^{d}k d^{d}k' |B_{kk'}|^2 \sim \int \frac{d^{d}k}{k^{d}}\ ,
 \eeq
which is logarithmically divergent, implying the absence of a unitary map between the corresponding Hilbert spaces.
 
The case $D=2$ provides an exception to this reasoning.  In this case, the stationarity condition $k_y + k_\mu' \partial_y\calx^\mu=0$ implies vanishing of the leading $\calo(k)$ behavior of the coefficient $C$, which does not generically vanish for $D>2$.  This follows from  the expression $\sqrt{q_f}n^\mu_f = (\partial_y \calx^1,\partial_y \calx^0)$.  This replaces equation \eqref{divsum} by one that is UV convergent, and allows a unitary map between Hilbert spaces.\footnote{For the large class of people who have developed their intuition studying evolution on two-dimensional geometries, this is one explanation for the possible intuition that evolution can in general be simply implemented by such a unitary map between Hilbert spaces.}

This problem clearly extends to slices in more general spacetimes, and a consequence of these statements is that the evolution from an initial slice $\Sigma_i$ to a final slice $\Sigma_f$ of a $D>2$ manifold $\calm$ cannot be described in terms of a unitary operator acting on a fixed Hilbert space defined as a representation of the basic field algebra \eqref{CCRy} on $\Sigma$.  Instead, the equivalence class of the complex structure $\{J\}$ defining the representation is generically different on the two slices.

This has implications for defining a Schr\"odinger picture in order to describe the evolving quantum state.  A na\"\i ve such picture doesn't exist on general spacetimes with $D>2$, since the evolution includes not only that of the quantum state, which might be described by a unitary operator $U(t)$, but also the evolution of the equivalence class of complex structures, $\{J\}_t$, which define the relation of the field operators $\phi(y)$, $\pi(y)$ to the Hilbert space of states.  

\section{Local analysis and Hadamard behavior}
\label{Localsec}

\subsection{Local description of quantum equivalence classes}
\label{Localeq}

The preceding discussion illustrates that one way to specify a complex structure and construct the Fock space of a theory is to begin with such a construction on an initial slice, and then infer the construction on a later slice by evolution of the corresponding solutions.  However, this approach would appear to have shortcomings.  First, there is a question of what, in general, would specify such an initial slice, and how the desired equivalence class would be specified on it. Secondly, in the case of an interacting theory, it is by no means simple to solve the equations of motion to find the quantum state at a later time.  

Both of these problems suggest the desirability of having a more local description of the construction of a physical equivalence class of complex structures, and of the corresponding Fock space.  In order to find this, note that since the underlying problem is a UV problem, we might investigate it by considering just the short distance behavior of fields and states, and specifically focussing on the behavior in the near vicinity of a point\footnote{This basic idea has also been formalized in the notions of local definiteness and  local stability of \cite{HNS,FrHa1,FrHa2}, and also in the formal analysis of \cite{Radz}.} $P\in \calm$.  For example, the integral \eqref{betaf} was analyzed by investigating specific large $k$ and $k'$ and finding a dominant contribution near one or more saddlepoints, but one might alternately start with the point $P$ and ask what modes dominantly contribute to $B_{kk'}$ from its vicinity; one could specifically consider for example wavepackets localized in a wider vicinity of $P$, but then look at the limit where these wavepackets are formed from very high momentum modes and investigate which of these dominantly contribute to $B_{kk'}$.  

For a fixed background metric, a sufficiently localized neighborhood of $P$ will be well-described as a neighborhood of flat space.  This can be exhibited by considering a local Lorentz frame $e^a_\mu$ on $\calm$,
\beq
ds^2= g_{\mu\nu} dx^\mu dx^\nu = e^a_\mu(x) e^b_\nu(x) \eta_{ab} dx^\mu dx^\nu\ .
\eeq
Then we can define local flat coordinates $X^a$ at $P$ via
\beq
dX^a= e^a_\mu(x_P) dx^\mu \quad \Rightarrow \quad X^a= e^a_\mu(x_P)(x^\mu-x^\mu_P)\ ;
\eeq
both choices are modulo an $SO(1,d)$ ambiguity corresponding to local Lorentz transformations.
Near $P$ and at sufficiently high frequency, a basis of classical solutions to \eqref{scalact} can be taken to have the flat space form
\beq\label{LHF}
 \varphi_k\simeq\caln_k \frac{e^{ik_a X^a}}{\sqrt{ 2 {\hat \omega_k}}} = \caln_k \frac{e^{ik_a e^a_\mu(x_P) (x^\mu -x^\mu_P)}}{\sqrt{  2{\hat \omega_k}}}\ 
\eeq
with $\caln_k$ a normalization factor, where $a=(\hat 0,\hat i)$ denote the frame indices, and  where $k_ak^a=-m^2$ gives
\beq
k_{\hat 0} =- \sqrt{ k_{\hat i}^2 + m^2}  = -{\hat \omega_k}\ ,
\eeq 
with sign corresponding to the positive frequency condition.

The spatial slice $\calx^\mu(t,y)$ that passes through $P$ will in general pass through it at some non-zero boost with respect to general coordinates $X^a$, and so will specify a particular Lorentz frame.  Consider an ADM parameterization of the metric, based on the foliation $\calx^\mu(t,y)$,
\beq
ds^2 = -N^2 dt^2 + q_{ij}(dy^i+N^i dt)(dy^j+N^j dt)\ .
\eeq
This defines the specific frame through
\beq
e^{\hat 0}_\mu dx^\mu = N dt\quad ,\quad e^{\hat i}_\mu dx^\mu = e^{\hat i}_j(dy^j + N^j dt)\ ,
\eeq
where $e^{\hat i}_j$ defines a frame corresponding to $q_{ij}$, $q_{ij}= \delta_{\hat k\hat l}e^{\hat k}_i e^{\hat l}_j$.  The local high-wavenumber, positive frequency solutions  \eqref{LHF} defined using this frame in a small neighborhood of $P$, with coordinate size $L$ and volume $V_P\simeq\sqrt{q(x_P)}L^{d}$, induce Cauchy data on the slice
\beq\label{CDH}
\gamma_k(y) = \frac{1}{\sqrt{ 2{\hat \omega_kV_P}}}(e^{ik_j y^j}, -i \sqrt{q_P} {\hat \omega_k }e^{ik_j y^j})\ ,
\eeq
with $q_P=q(x_P)$, defining a complex structure.  The normalization here is chosen assuming that modes are restricted to $V_P$, and so have norm $\langle\gamma_k^*,\gamma_{k'}\rangle =\delta_{kk'}$; for more discussion of normalization see Appendix~\ref{appa}.\footnote{An approach to a careful definition of such orthogonal modes is to use a partition of unity to restrict to respective regions. Appendix~\ref{appa} outlines such a setup, though a fully precise treatment is left for future work.}    Notice that in terms of the covariant momentum components $k_i$, 
\beq\label{omegakhat}
 {\hat \omega_k} = \sqrt{q^{ij}(x_P) k_ik_j + m^2} \ .
 \eeq

Suppose that instead we worked with a fixed (slice-independent) complex structure; for example, if the manifold and slices were flat in the past, this might be the corresponding standard flat complex structure considered in Sec.~\ref{EvolEC}.  The associated positive frequency solutions correspond to Cauchy data
\beq\label{CDl}
\bar\gamma_k=\frac{1}{\sqrt{ 2\omega_k L^d}} (e^{ik_j y^j}, - i \omega_k e^{ik_j y^j}) \ ,
\eeq
with $\omega_k$ given by \eqref{omegak}.
For a given wavenumber $k_i$, comparison shows a difference between  complex structures with 
\beq\label{CS}
\tau_k= i\sqrt{q_P}{\hat \omega_k} = i\sqrt{q_P}\sqrt{q^{ij}(x_P) k_ik_j + m^2}
\eeq
for the local positive frequency data \eqref{CDH}, and 
and $\bar\tau_k=i\omega_k$ for the data \eqref{CDl} corresponding to the fixed complex structure.  
Comparing \eqref{CDl} with \eqref{CDH} shows that the fixed complex structure provides Cauchy data for a local solution
\beq\label{fixsoln}
\bar \varphi_k = A_k \varphi_k + B_k \varphi_{-k}^*\ ,
\eeq
where we find
\beq\label{ABc}
A_k=\hf\left( \sqrt{\frac{{\sqrt{q_P}}{\hat \omega_k}}{\omega_k}} + \sqrt{\frac{\omega_{k}}{\sqrt{q_P}{\hat \omega_k}}} \right)\quad ,\quad B_k= \hf\left( \sqrt{\frac{\sqrt{q_P}{\hat \omega_k}}{\omega_k}} -\sqrt{\frac{\omega_{k}}{\sqrt{q_P}{\hat \omega_k}}} \right)\ .
\eeq
Thus, in this local limit the Bogolubov coefficients are
\beq\label{ABloc}
A_{kk'}=\langle\gamma_k^*,\bar \gamma_{k'}\rangle \simeq A_k \delta_{kk'}\quad , \quad B_{kk'}=\langle\gamma_k,\bar \gamma_{k'}\rangle\simeq B_k\delta_{k,-k'}\ .
\eeq

Again, the nonvanishing $B_{kk'}$ leads to a transformation that is not Hilbert-Schmidt and so cannot be unitarily implemented.  The na\"ive high-energy expression $\sum_k |B_{k}|^2$ appears to have a higher-order divergence than found in Sec.~\ref{EvolEC}.  However, this is partly an artifact of the local approximation.  The full solutions only take plane wave form \eqref{LHF} near $P$, and depart from this outside of a small neighborhood, so the integral of products of plane waves is correspondingly restricted in volume.
For example, in the previous case of a nontrivial slice in flat space, there were important departures from plane wave form on scales $\Delta y\sim 1/\sqrt{k \partial^2 \calx}$, seen from \eqref{slicedat},
and so in $B_{kk'}$ the effective volume of the integral with plane wave behavior was $V\sim (\Delta y)^d \sim 1/(k\partial^2\calx)^{d/2}$. 
This leads to the additional factors $\sim 1/k^{d/2}$ found there, and thus to the logarithmic divergence found in \eqref{divsum}.  One expects similar suppressions in the case of a curved background. 

One can see from this that the physical equivalence class is determined by the complex structure \eqref{CS} 
associated to local Cauchy data  \eqref{CDH}.  A necessary condition for this complex structure to stay in the equivalence class of the fixed complex structure is vanishing of the $\calo(k^0)$  term of an expansion of $B_k$ in \eqref{ABc} in powers of $1/k$. More generally, a necessary condition for complex structures associated with two different slices to be in the same equivalence class is agreement of their corresponding $\tau_k$'s of \eqref{CS} to leading order in large $k$.  This condition extends the  preceding discussion to more general spacetimes, where one does not have evolution from a flat slice of flat space in the past.  
If this is not satisfied, there is an infinite difference of particle number \eqref{Bsquared} between the states, indicating the lack of a Hilbert-Schmidt transformation between them.  We leave the careful investigation of sufficiency for equivalence, which in flat space depended on the $\sim 1/k^{d/2}$ factors, for future work.

\subsection{Connection to Hadamard behavior}

We can also see that the preceding condition is related to the Hadamard condition for behavior of the two point function\cite{Hada,Waldbook}, by examining the latter's behavior.  Again working in a small neighborhood of a point $P$, consider expanding the field in either a good local basis $\varphi_k(x)$, as in \eqref{LHF}, or a ``bad" basis $\bar \varphi_k(x)$,
\beq
\varphi(x)= \sum_k \left(  a_k \varphi_k +  a_k^\dagger \varphi_k^*\right) = \sum_k \left( \bar a_k \bar \varphi_k + \bar a_k^\dagger \bar \varphi_k^*\right) \ .
\eeq
If the Hilbert space is defined using the good complex structure, then from \eqref{LHF}, \eqref{CDH}
\beq\label{flatapprox}
\langle 0| \varphi(x) \varphi(x') | 0\rangle = \sum_k  \varphi_k(x)  \varphi_k^*(x') \simeq \int \frac{d^dk_{\hat i}}{(2\pi)^d 2 {\hat \omega_k}} e^{ik_a(X^a-X^{a\prime})}\ ,
\eeq
where we convert the sum to an integral as in \eqref{infvol}, and transform from components $k_i$ to $k_{\hat i}$.

This two-point function thus has local behavior corresponding to that of the flat space propagator, and this gives the expected leading Hadamard singular behavior. Specifically, the calculation of the flat space expression \eqref{flatapprox}, for example with $d=3$, can be found in field theory references such as  \cite{weinberg1995quantum}, which for spacelike separation gives
\bea\label{flatprop}
\langle \bar 0| \varphi(X) \varphi(X') | \bar 0\rangle &=& \frac{m}{4\pi^2}\frac{1}{\sqrt{\Delta X_i^2-\Delta T^2}}K_1\left(m \sqrt{\Delta X_i^2-\Delta T^2}\right)\ .
\eea
The expansion of this in the local limit $\sigma=\frac{1}{2}(\Delta X_i^2-\Delta T^2) \to 0$ then gives an expression of the general form 
\beq\label{Haddef}
\langle 0| \varphi(x) \varphi(x') | 0\rangle \sim \frac{U(x, x')}{8\pi^2\sigma}+V(x, x')\ln{\sigma}+W(x, x')\ ,
\eeq
where $V(x, x')$ and $W(x, x')$ are series involving positive integer powers of $\sigma$, and $U(x, x) =1$.   
When generalized to curved backgrounds, \eqref{Haddef} defines the Hadamard form of the expansion of the two-point function, with $2\sigma$ the geodesic distance between $x$ and $x'$;
there are further conditions on $U$, $V$, and $W$, and a discussion of these and the generalization to arbitrary $d$ appears in  \cite{decanini2008hadamard}.

Physically acceptable states must give such  a Hadamard expansion, or in other words, the short-distance behavior of the two-point function must match to the behavior of the propagator in Minkowski space. This local behavior being in agreement with  that of the vacuum state is in particular required to correctly define a renormalized stress-energy tensor in a curved spacetime, as the leading singular behavior comes from its terms  quadratic in the fields and their derivatives, and so it has singular behavior related to that of the two-point function. The short-distance behavior of each is dominated by the high-energy limit of the field, and requiring it to be nearly that of Minkowski space then allows the definition of the appropriately renormalized vacuum energy and stress tensor~\cite{wald1995quantum}.

On the other hand, using the vacuum $|\bar 0\rangle$ corresponding to the bad (fixed) complex structure, ${\bar a_k}|\bar 0\rangle=0$, gives, from \eqref{fixsoln}
\bea
\langle \bar 0| \varphi(x) \varphi(x') | \bar 0\rangle 
&=& \sum_k  \bar \varphi_k(x)  \bar \varphi_k^*(x') \cr &\simeq &
\sum_k \Big[ |A_k|^2 \varphi_k(x)  \varphi_k^*(x') + |B_k|^2 \varphi_{-k}^*(x)  \varphi_{-k}(x')\cr &&+ A_kB_k^*\varphi_k(x)  \varphi_{-k}(x') + A_k^*B_k\varphi_{-k}^*(x)  \varphi_{k}^*(x') \Big] \ .\label{twopt}
\eea
Then the two-point function arising from the bad complex structure can be rewritten using the  Bogolubov coefficients \eqref{ABc} and the preceding replacement of the sum over modes by an integral.  Note that each term has identical dependence on the spatial distance $\Delta X^i = X^i -X'^i$, but has differing dependence on the times denoted by $T=X^0$ and $T'=X^{0\prime}$:
\begin{align}
\langle \bar 0| \varphi(x) \varphi(x') | \bar 0\rangle 
\simeq& \int \frac{d^dk_{\hat i}}{(2\pi)^d 2 {\hat \omega_k}} \frac{1}{4} \left\{\left(\frac{\sqrt{q_P}\hat \omega_k}{\omega_k}+2+\frac{\omega_k}{\sqrt{q_P}\hat \omega_k}\right)e^{-i \hat \omega_k \Delta T}  \right. 
+\left(\frac{\sqrt{q_P}\hat \omega_k}{\omega_k}-2+\frac{\omega_k}{\sqrt{q_P}\hat \omega_k}\right)e^{i \hat \omega_k \Delta T} \cr 
&\quad+\left(\frac{\sqrt{q_P}\hat \omega_k}{\omega_k}-\frac{\omega_k}{\sqrt{q_P}\hat \omega_k}\right)
\left.\left[e^{-i \hat \omega_k (T+T')}+
e^{i \hat \omega_k (T+T' )} \right] \right\}e^{i k_{\hat i}\Delta X^{ i}}\cr
=&\, \hf \int \frac{d^dk_{\hat i}}{(2\pi)^d 2 {\hat \omega_k}} 
e^{i k_{\hat i}\Delta X^{ i}}
\left\{ \left( \frac{\sqrt{q_P} \hat \omega_k }{\omega_k} +\frac{\omega_k}{\sqrt{q_P} \hat \omega_k}\right)
\cos(\hat \omega_k\Delta T)
- 2i\sin(\hat \omega_k\Delta T)\right. \cr 
&\left.+\left(\frac{\sqrt{q_P}\hat \omega_k}{\omega_k}-\frac{\omega_k}{\sqrt{q_P}\hat \omega_k}\right)\cos\left[\hat \omega_k(T+T')\right]\right\}\ .
\label{twoptexact}
\end{align}

This expression for the two-point function generically has singular behavior that is not Hadamard.  In this expression, we find
\beq
\omega_k=\sqrt{{\tilde q}_P^{\hat i\hat j}k_{\hat i}k_{\hat j}+ m^2}
\eeq
where
\beq
{\tilde q}^{\hat i\hat j}= e^{\hat i}_k e^{\hat j}_k\ .
\eeq
The resulting expression is not generically rotation invariant, making the generic singular structure difficult to exhibit by evaluating the integral.  In addition, notice that the two-point function is not time-translation invariant.  The leading singular behavior is expected to come from the leading terms in an expansion of the coefficients in \eqref{twoptexact} in $1/\hat k^2$.  The non time-translation invariant term has coefficient 
\beq\label{ABeq}
A_kB_k^*= \frac{1}{4}\left(\frac{\sqrt{q_P}\hat \omega_k}{\omega_k}-\frac{\omega_k}{\sqrt{q_P}\hat \omega_k}\right) = \frac{1}{4}\left(\frac{\sqrt {q_P \hat k^2}}{\sqrt{{\tilde q}^{\hat i\hat j}_P k_{\hat i}k_{\hat j}}}- \frac{\sqrt{{\tilde q}^{\hat i\hat j}_P k_{\hat i}k_{\hat j}}}{\sqrt{ q_P\hat k^2}} \right) +\calo\left(\frac{1}{\hat k^2}\right)\ ,
\eeq
and so even the leading singularity is in general difficult to evaluate due to the rotational non-invariance.  

However, the non-Hadamard behavior of this two-point function may be illustrated in the special case where the slice metric is locally isotropic, in the neighborhood of $P$.  Specifically, if
\beq
e_{jP}^{\hat i}=a\delta_j^{\hat i}\ ,
\eeq
then ${\tilde q}_P^{\hat i\hat j} = a^2 \delta^{\hat i\hat j}$, 
and \eqref{ABeq} becomes
\beq
A_kB_k^*=\frac{1}{4}\left( \frac{a^d\sqrt{\hat k^2+m^2}}{\sqrt {a^2 \hat k^2 + m^2}}- \frac{\sqrt {a^2 \hat k^2 + m^2}}{a^d\sqrt{\hat k^2+m^2}} \right)
=\frac{1}{4}\left(a^{d-1}-\frac{1}{a^{d-1}}\right) + \calo\left(\frac{1}{\hat k^2}\right)\ .
\eeq
Notice here the special status of $d=1$, described earlier. 
 The two-point function \eqref{twoptexact} then becomes 
\bea
\langle \bar 0| \varphi(x) \varphi(x') | \bar 0\rangle &=& \int \frac{d^dk_{\hat i}}{(2\pi)^d 2 {\hat \omega_k}} \frac{1}{2} e^{i k_{\hat i}\Delta X^i}\left[\left(a^{d-1}+\frac{1}{a^{d-1}}\right) \cos(\hat \omega_k\Delta T)-2i \sin(\hat \omega_k\Delta T) \right.\cr &&\left.+\left(a^{d-1}-\frac{1}{a^{d-1}}\right) \cos(\hat\omega_k(T+T')) + \calo\left(\frac{1}{\hat k^2}\right)
\right]\ .\label{twoptapprox}
\eea
The leading singular behavior may be evaluated by first integrating over the angular variables, and then evaluating the remaining integral over $k$ in terms of modified Bessel functions.\footnote{Often two-point functions in Minkowski space are calculated by going into a reference frame which simplifies these integrals.  While the integrals here are not Lorentz invariant, they may be directly related to the Lorentz-invariant expression \eqref{flatapprox} by taking real and imaginary parts, or by substitution of the time argument.}
For example with spacelike separated points, and $d=3$, 
\bea\label{nonHS}
\langle \bar 0| \varphi(x) \varphi(x') | \bar 0\rangle &\simeq& \frac{m}{8\pi^2}\left(a^{2}+\frac{1}{a^{2}}\right)\frac{1}{\sqrt{\Delta X_i^2-\Delta T^2}}K_1\left(m \sqrt{\Delta X_i^2-\Delta T^2}\right) \cr
&+&\frac{m}{8\pi^2}\left(a^{2}-\frac{1}{a^{2}}\right)\frac{1}{\sqrt{\Delta X_i^2-(T+T')^2}}K_1\left(m \sqrt{\Delta X_i^2-(T+T')^2}\right)\ .
\eea
The first term has a singularity similar to Hadamard form, but with a different coefficient.  The singularity in the second term, at ${(T+T')^2=\Delta X_i^2}$ is not of Hadamard form.\footnote{This statement refers to failure to be globally Hadamard, as opposed to locally.  See \cite{Gonnella:1989np, Radzikowski, kay1991theorems} for further clarifications. The inclusion of such a spacelike singularity was conjectured to be disallowed for Hadamard states in \cite{Gonnella:1989np}, which explored several examples and found that states with such singularities  do not correspond to quasi-free states. The conjecture was proven in the thesis of Radzikowski \cite{Radzikowski} for globally hyperbolic spacetimes. The paper \cite{kay1991theorems} also introduces a definition of globally Hadamard as a requirement for physical states that is partially motivated by the exclusion of nonlocal singularities.}

Such leading-order contributions to non-Hadamard singularities from \eqref{twoptexact} are eliminated when the Bogolubov coefficients $B_{kk'}$ of \eqref{ABloc}, \eqref{ABc} vanish at $\calo(k^0)$.  When that is true, the leading behavior of eq.~\eqref{twoptexact} reduces to the flat propagator \eqref{flatprop}, of Hadamard form.  
This $\calo(k^0)$ vanishing condition is the same as previously 
discussed for the transformation between the complex structures to be Hilbert-Schmidt, that is for them to lie in the same equivalence class.  
In this sense, we see a direct connection between the Hadamard behavior of the two-point function, and the condition determining the equivalence class of the complex structure.

This discussion thus directly illuminates the connection between the condition determining the physical equivalence class of complex structures and the Hadamard condition on the two-point function.  In doing so, it complements more abstract arguments previously given in the literature\cite{Wald1979,FNW}.  Specifically, consider spacetimes with compact spatial sections to eliminate IR divergences.  In \cite{FSW}, it is shown that if in a free theory the Hadamard condition is satisfied in the neighborhood of an initial slice, it holds throughout the spacetime, for sufficiently smooth spacetimes (Ref.~\cite{FSW} assumes $C^\infty$, although this condition can  be relaxed as is further described below).   If the spacetime has static regions in the past and future, their metrics naturally determine complex structures in the respective regions.  If the distributions $F_1(x,x')$ and $F_2(x,x')$ denote the corresponding vacuum two-point functions, then $S=F_1-F_2$ will be finite, due to cancellation of the Hadamard singularities.  
Since  $S$ is a difference of vacuum two-point functions, it is also a solution to the Klein-Gordon equation in both arguments $x$, $x'$.  
Then, $S$ may be split into  positive and negative frequency components in each variable, and related to the Bogolubov coefficients $A$ and $B$ connecting the past and future bases. The smoothness of $S$ via the cancellation of singularities is shown\cite{Wald1979,FNW} to imply the Hilbert-Schmidt condition
\beq
\label{Conv2}
\sum_{AB}|B_{AB}|^2 <\infty \ ,
\eeq
for the Bogolubov transformation relating the past and future bases, implying finiteness of particle number and existence of an S-matrix in this context.  
Moreover, these arguments can be extended to more general spacetimes by a deformation argument from such a spacetime to one with static past and future regions\cite{FNW}.

\section{Some examples}

\subsection{Cosmological evolution}

Cosmological spacetimes give particularly simple curved space examples of the 
the preceding issues with defining a unitary map between Hilbert spaces at different times in the absence of time-translation symmetry, and the relation to Hadamard behavior.   As in general, there are both UV and IR issues with defining the states and Hilbert space.  The IR issues, which are discussed for example in  \cite{Waldbook,Wald2009,WittQFT}, can be controlled, either by introducing an appropriate set of boundary conditions at infinity, leading to unitarily equivalent representations, or working in a closed universe.  That leaves the UV issue of infinite changes in particle number, associated with the inequivalence of complex structures on different slices.

As previously, we will choose the option of working in a closed spacetime, {\it e.g.} with periodic spatial identification $x^i\simeq x^i+L$,
and consider a FLRW spacetime metric
\begin{equation}\label{Cosmetric}
    ds^2 = a^2(\eta)(-d\eta^2+dx^{i2})\ .
\end{equation}
UV unitarity for such spacetimes has been studied in IIC of \cite{AgAs}, and we follow part of their presentation.  Specifically, consider the case where the scale factor transitions between constant values in the distant past and future, $a(\eta_-) = a_-$ and $a(\eta_+) = a_+$, respectively, so  the spacetime is stationary on the initial and final slices.  Modes which are positive frequency in the stationary past and future are defined according to the associated timelike Killing vector on those slices, and are associated with Cauchy data (compare \eqref{flatdata})
\begin{align}
  \gamma_{k}^-:\quad  \phi_k^-(\vec{x})&= \frac{1}{\sqrt{2\omega_{k}^- a_-^{d-1} L^d}}e^{ i \vec{k} \cdot \vec{x}}\quad ,\quad \pi_k^-( \vec{x})=- i \sqrt{\frac{\omega_{k}^- a_-^{d-1}}{2L^d}} e^{i \vec{k} \cdot \vec{x}} \ ,\label{initialmode}\\
   \gamma_{k}^+:\quad  \phi_k^+(\vec{x})&= \frac{1}{\sqrt{2\omega_{k}^+ a_+^{d-1} L^d}}e^{ i \vec{k} \cdot \vec{x}}\quad ,\quad \pi_k^+( \vec{x})=- i \sqrt{\frac{\omega_{k}^+ a_+^{d-1}}{2L^d}} e^{i \vec{k} \cdot \vec{x}}\ ,\label{finalmode}
\end{align}
respectively, recalling that $\pi$ is a scalar density, and with $\omega_{k}^{\pm} = \sqrt{k^2+m^2a^2_{\pm}}$. 

The Bogolubov coefficients
\beq 
A_{kk'}=\langle\gamma_k^{+*}, \gamma_{k'}^-\rangle= A_k \delta_{kk'}\quad , \quad B_{kk'}=\langle\gamma_k^{+},\gamma_{k'}^{-}\rangle= B_k\delta_{k,-k'}\
\eeq
relating these bases can be calculated, giving\footnote{The overall sign for $B_{kk'}$ differs from \cite{AgAs} due to the chosen convention.}
\beq
   A_{k}=\hf\left[ \sqrt{\frac{\omega^-}{\omega^+}}\left(\frac{a_-}{a_+}\right)^{\frac{d-1}{2}}+\sqrt{\frac{\omega^+}{\omega^-}}\left(\frac{a_+}{a_-}\right)^{\frac{d-1}{2}}\right]\quad ,\quad
   B_{k}= \hf\left[ \sqrt{\frac{\omega^-}{\omega^+}}\left(\frac{a_-}{a_+}\right)^{\frac{d-1}{2}}-\sqrt{\frac{\omega^+}{\omega^-}}\left(\frac{a_+}{a_-}\right)^{\frac{d-1}{2}}\right]\ \label{divbog}
\eeq
(compare \eqref{ABc}, \eqref{ABloc}).
To leading order in large $k$, the expressions in the brackets are constant, so the first term in the expansion of $B_{kk'}$ is a constant independent of $k$. From the sum \eqref{divsum} we see that this transformation is clearly not Hilbert-Schmidt.  This means that the bases $\gamma_k^-$ and $\gamma_k^+$ correspond to different equivalence classes of complex structures, {\it i.e.}
define representations of the commutation relations \eqref{CCRs} that are not related by a unitary transformation. 
This  gives a simple example of the trouble with a na\"ive definition of unitary evolution in a  Schr{\"o}dinger picture in a time dependent spacetime.

Lack of Hadamard behavior  also serves here as a diagnostic for this inequivalence.  Analogously to  \eqref{twopt}, the two-point function in the $\gamma_k^-$ representation, with vacuum state $|0^-\rangle$, can be calculated for points $x,x'$ in the future region, giving ({\it e.g.} with $d=3$)
\bea\label{tpm}
\langle 0^-| \varphi(x) \varphi(x') |0^-\rangle 
&=& \sum_k \Big[|A_k|^2 \varphi_k(x)  \varphi_k^*(x') + |B_k|^2 \varphi_{-k}^*(x)  \varphi_{-k}(x')\cr &&+ A_k B_k^*\varphi_k(x)  \varphi_{-k}(x') + A_k^*B_k\varphi_{-k}^*(x)  \varphi_{k}^*(x') \Big] \cr
&\simeq& \int \frac{d^3k}{(2\pi)^3 2 \omega_{k}} \frac{1}{a_+^2}e^{i \vec{k}\cdot\Delta \vec{x}}(|A_k|^2 e^{-i\omega_k(\eta-\eta')} + |B_k|^2 e^{i\omega_k(\eta-\eta')}\cr &&+ A_k B_k^*e^{-i\omega_k(\eta+\eta')}+ A_k^*B_ke^{i\omega_k(\eta+\eta')}) \ ,
\eea
where here $\varphi_k=\varphi_k^+$, $\omega_k=\omega_k^+$.  Here due to angular independence of $A_k$ and $B_k$, the angular integral may be performed, but the remaining integral over $k$ is nontrivial due to the dependence of $A_k$ and $B_k$ on $\omega_k^\pm$.  However, again the leading singular behavior follows from an expansion of the coefficients in $1/k$, 
\beq
A_k = \hf\left[  \left(\frac{a_-}{a_+}\right)^{\frac{d-1}{2}}+ \left(\frac{a_+}{a_-}\right)^{\frac{d-1}{2}}\right] +\calo\left(\frac{1}{k^2}\right)
\quad ,\quad
   B_{k}= \hf \left[\left(\frac{a_-}{a_+}\right)^{\frac{d-1}{2}}-\left(\frac{a_+}{a_-}\right)^{\frac{d-1}{2}}\right]+\calo\left(\frac{1}{k^2}\right)\ .
\eeq
Similarly to \eqref{nonHS}, \eqref{tpm} then gives for $d=3$
\begin{align}
\langle0^-| \varphi(x) \varphi(x') | 0^-\rangle\simeq& \frac{m}{8\pi^2}\frac{1}{a_+^3a_-^2}(a_+^4+a_-^4)\frac{1}{\sqrt{\Delta x_i^2-\Delta \eta^2}}K_1\left(m a_+\sqrt{\Delta x_i^2-\Delta \eta^2}\right) \cr
+&\frac{m}{8\pi^2}\frac{1}{a_+^3a_-^2}(a_+^4-a_-^4)\frac{1}{\sqrt{\Delta x_i^2-(\eta+\eta')^2}}K_1\left(ma_+ \sqrt{\Delta x_i^2-(\eta+\eta')^2}\right)\ .
\end{align}
The first term has a leading singularity of Hadamard form, \eqref{Haddef}, albeit with a nonstandard coefficient.
The leading non-vanishing of $B_k$, responsible for violating the Hilbert-Schmidt condition, also produces the non-Hadamard singular form of the second term.

Although the Hilbert spaces $\calh^-$ and $\calh^+$ in the past and future correspond to inequivalent representations of the commutation relations \eqref{CCRs}, there is a unitary map between them, generated by the evolution, for sufficiently differentiable metrics.  This is related to a definition of an S-matrix\cite{wald1995quantum}.
This is found by starting with modes that are positive frequency in the past, as with  \eqref{initialmode}.  The evolution of this initial data through the expanding spacetime defines mode solutions $\varphi_k^-(\eta, \vec x)$.  Likewise, mode solutions $\varphi_k^+(\eta, \vec x)$ may be defined that have positive frequency Cauchy data \eqref{finalmode} in the far future.  The symplectic product \eqref{prodsoln} then defines Bogolubov coefficients
\beq\label{transf}
\alpha_{kk'}=i\Omega(\varphi_k^{+*},\varphi_k^-)\quad ,\quad \beta_{kk'}=i\Omega(\varphi_k^{+},\varphi_k^{-})\ .
\eeq
Here the inner products are defined on a single Cauchy slice, rather than being between data defined on different such slices, {\it e.g} in past and future.  These inner products are independent of this slice, {\it i.e.} are time independent.
These coefficients do respect the Hilbert-Schmidt condition,
\begin{equation}\label{Convergence}
    \sum_{kk'} |\beta_{kk'}|^2 <\infty \ ,
\end{equation}
{\it e.g.} where the metric is assumed to be $\mathcal{C}^{\infty}$ \cite{FNW}, though this condition may be relaxed to a weaker continuity condition.  There is a corresponding transformation between the creation and annihilation operators $b_k^\dagger$, $b_k$ associated with the respective modes, 
\begin{equation}\label{Stransf}
    b^{+}_k = \sum_{k'}i \big[\Omega(\varphi_k^{+*},\varphi_{k'}^-)b^{-}_{k'}+\Omega(\varphi_k^{+*},\varphi_{k'}^{-*})b^{-\dagger}_{k'}\big]
\end{equation}
And, any past state $|\psi^-\rangle\in \calh^-$ may be mapped to a future state $|\psi^+\rangle\in \calh^+$ by the corresponding unitary transformation.

Examples of derivation of the Bogolubov coefficients \eqref{transf} can be given in special cases\cite{PaTo,AgAs}.
In fact, infinite differentiability can be relaxed to  $C^2$  as can be seen by using adiabatic methods.\footnote{For discussion and extensive references, see \cite{PaTo}.}
In a cosmological spacetime \eqref{Cosmetric}, the modes on the initial slice \eqref{initialmode} evolve according to the equations of motion, 
which for the example of a massive scalar in $D=4$ become
\beq
-\varphi'' -2\frac{a'}{a}\varphi' + \partial_i^2\varphi = m^2 a^2\varphi
\eeq
with prime denoting derivative with respect to $\eta$.  This has solutions of the form
\beq
\frac{\chi_k(\eta)}{a(\eta)} e^{i\vec k\cdot \vec x}\ ,
\eeq
with
\beq
\chi_k'' +\left(k^2 + m^2 a^2 -\frac{a''}{a}\right)\chi_k=0\ .
\eeq
For modes with high frequencies as compared to the inverse timescale for change in $a$, this can be solved in an adiabatic
expansion using the WKB-like ansatz
\begin{equation}\label{WKBmode}
   \varphi_k^-(\eta, \vec{x})= \frac{1}{a(\eta)\sqrt{2W(\eta)}}e^{-i\int^{\eta}_{\eta_-} W(\tau) d\tau}e^{ i \vec{k} \cdot \vec{x}} \ ,
\end{equation}
with which the equation of motion becomes
\begin{equation}\label{adexp}
    W^2 = k^2 {+ m^2 a^2} +W^{1/2}\frac{d^2}{d\eta^2}W^{-1/2} -\frac{a''}{a}\ .
\end{equation}
This is then solved order by order in $1/\omega(a)$, with $\omega(a)=\sqrt{k^2+m^2a^2}$, through an expansion $W(\eta) = \sum_i W_i$, in which the odd terms are found to be zero.  The solutions that are positive frequency in the past are given by
\beq\label{wexp}
W=\omega(a) + \frac{1}{2\omega(a)}\left\{\omega^{1/2}(a)\partial_\eta^2\left[ \omega^{-1/2}(a)\right] -\frac{a''}{a}\right\} + \calo\left(\frac{1}{\omega^3(a)}\right)
\eeq
One can then readily compute the Bogolubov transformation between these solutions, and a basis of solutions that are defined to be positive frequency at a later time.

For simplicity, consider the case where $a$ transitions from an initial regime with constant $a_-$ to a final one with constant $a_+$.  In this case, these approximate solutions  are clearly also positive frequency in the constant $a_+$ regime, due to the vanishing of the derivative terms in \eqref{wexp}, and so the resulting  $\beta_{kk'}$ vanish in an expansion in $1/\omega$.  This is consistent with the statement that the exact coefficients are exponentially suppressed at large $k$\cite{Kuls,PaTo}.  As a result, the transformation is clearly Hilbert-Schmidt, giving an example of the unitary transformation defined by  \eqref{Stransf}, and this is also expected to extend to the non-constant regime in a more thorough analysis.

We can likewise examine the behavior of the correlation functions.  The two-point function is 
\beq
\langle 0^-| \varphi(x) \varphi(x') |0^-\rangle 
= \sum_k \varphi_k^-(x)  \varphi_k^{-*}(x') \ .
\eeq
In the constant $a_+$ region, the approximate solutions \eqref{WKBmode} become the corresponding positive-frequency solutions, giving the Hadamard result 
 \beq
 \langle 0^-| \varphi(x) \varphi(x') | 0^-\rangle \simeq \int \frac{d^3k}{(2\pi)^3} \frac{1}{a_+^2}\frac{1}{2\omega}e^{-i\omega(\eta-\eta') +i \vec{k}\cdot (\vec{x}-\vec{x'})} = 
\frac{m}{4\pi^2}
\frac{1}{a_+}
\frac{1}{\sqrt{\Delta x_i^2-\Delta \eta^2}}
K_1 (m a_+\sqrt{\Delta x_i^2-\Delta \eta^2})\ ,
\eeq
where the last equality follows in the case of spacelike separation.  The corrections to the approximate solutions  \eqref{WKBmode} are exponentially suppressed at large $k$, and so do not modify this singular behavior.

Such cosmological examples therefore clearly illustrate the principles in action.  If we choose a good complex structure at an initial time, that will generically not be in the equivalence class of a good complex structure at a later time,  resulting in the non-unitary transformation given by \eqref{divbog}. However, the complex structure
that follows from the {\it evolution} of a good initial complex structure is in the same equivalence class as a good final complex structure.  These statements together imply that the  time evolution takes the initial state out of the equivalence class of an initial ``fixed" complex structure.  This is a problem for a standard Schr\"odinger picture description.
And, corresponding to the second statement, the initial Hadamard two-point function evolves to a
final Hadamard two-point function.
However, this discussion clearly depends on the solution of the non-interacting equations of motion, and does not obviously extend to an interacting theory.  Further discussion of interacting theories, and the relevance of the local description of the physical equivalence classes, will be briefly discussed below.

\subsection{Black hole evolution}

Another important example of our general discussion is that of black hole spacetimes.  For simplicity, we focus on the Schwarzschild case, with metric given in ingoing Eddington-Finkelstein coordinates $(x^+, r, \Omega)$,
\beq
ds^2= -f(r) dx^{+2} + 2dx^+ dr + r^2 d\Omega^2\ ,
\eeq
where $f=[1-(R/r)^{d-2}]$.  Derivations of Hawking radiation have traditionally been given in an S-matrix approach\cite{Hawk,WaldHR}, relating modes in the asymptotic future and past.  However, it has been widely thought that we can describe evolution of the quantum state on a spatial slicing of the geometry, and aspects of a concrete description of such evolution have for example been explored in \cite{SEHS,GiPe1,GiPe2}.  In the limit where backreaction is neglected, the spacetime is usefully described in a ``stationary slicing"\cite{NVU,SEHS},
defined by
\beq\label{slicedef}
x^+=t+S(r)
\eeq
with slice function $S(r)$.  An initial such slice $\Sigma$, as illustrated in Fig.~1,  is translated forward by the Killing vector.

\begin{figure}[h]
 	\begin{center}
 		\includegraphics[width=0.70\textwidth]{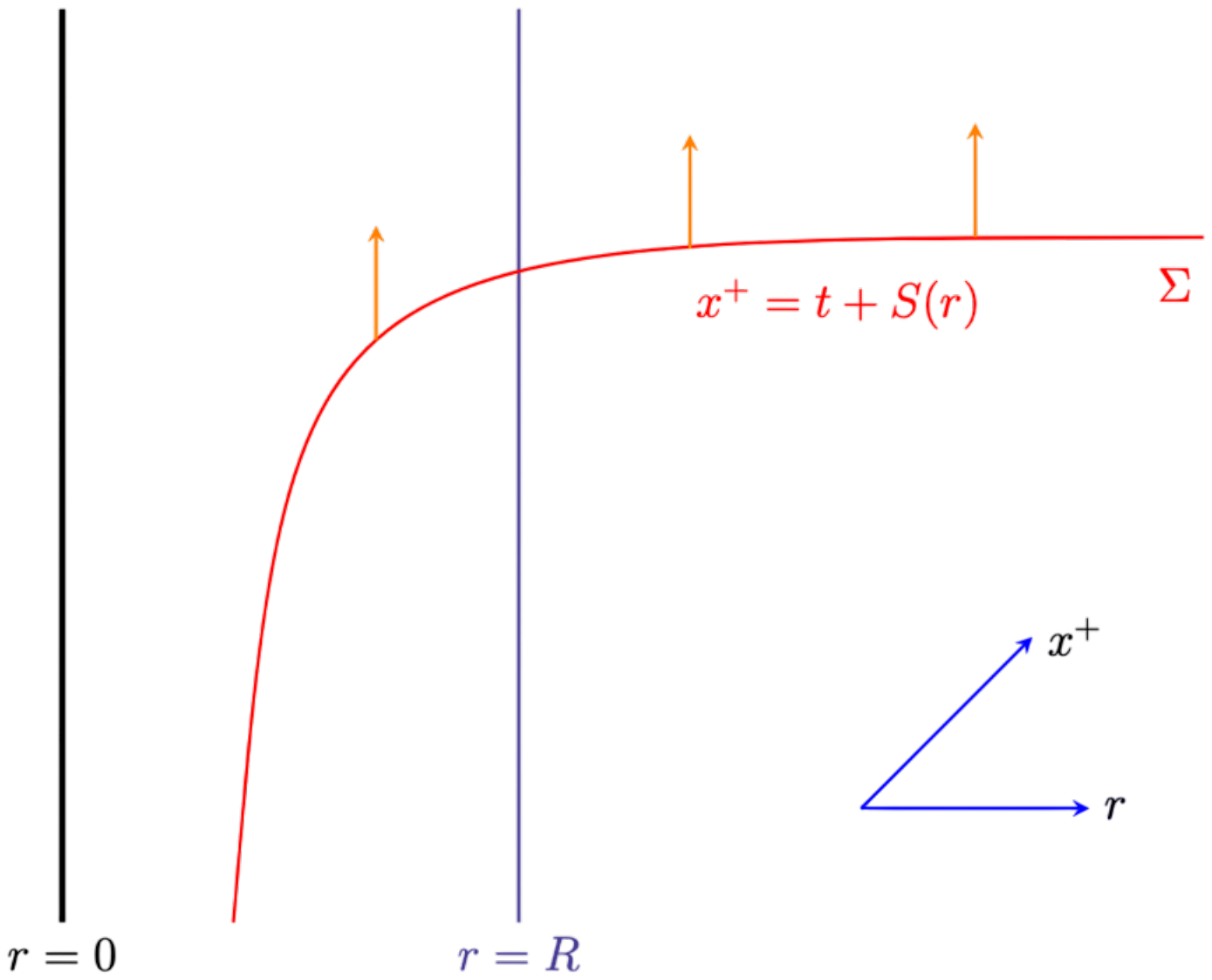} 
 		\caption{Construction of a stationary slicing of a Schwarzschild black hole spacetime, pictured in Eddington-Finkelstein coordinates, with horizon at $r=R$.  A slice $\Sigma$ with shape determined by $S(r)$ is translated by the Killing vector to give the foliation.}
 		\label{fig-slice}
 	\end{center}
 \end{figure}

On the slice $\Sigma$, we expect to be able to choose symplectic bases $\phi_I$, $\pi_{I'}$, as described in Sec.~\ref{CanonQ}, and then to be able to choose a complex structure $J$ in such a basis, determining the positive-frequency subspace.

One approach to this, studied in \cite{GiPe1}, is to consider an eigenbasis of time (here, $x^+$) translation.  As discussed there, one can find such energy eigenmodes, 
$\tilde \varphi_{\omega lm}$, that are regular at the horizon, $r=R$.  However, these ``in" modes are purely ingoing at the horizon, and so do not form a complete basis.  
The energy eigenmodes that are outgoing at the horizon, namely the ``up" modes $\varphi_{\omega lm}$ and the ``inside" modes $\hat \varphi_{\omega lm}$, are also singular there.  As a result the two-point function in the corresponding na\"\i ve vacuum, with positive frequency definition given by this basis, will be singular (non-Hadamard) at the horizon.

To describe a quantum state that is regular at the horizon, we would like to find a set of regular modes that give a complete basis.  In principle, one expects there to be a space of Cauchy data complementary to the regular in modes,  corresponding to ``regular outgoing" solutions $\varphi_{klm}$, that complete the basis and are orthogonal to the in basis; some criteria for determining these solutions were discussed in \cite{GiPe1}.\footnote{Explicit examples in two dimensions were given in \cite{SEHS,SE2d}.}  However, we so far lack an explicit construction of such modes; difficulties arise from the nontrivial reflection/transmission coefficients for $D>2$, combined with the orthogonality condition.

In the absence of such explicit constructions, one can characterize general aspects of the modes and quantization.  Specifically, we expect bases $(\phi^a_{klm},\pi^a_{klm})$ for  Cauchy data to exist, where $k$ is a continuous quantum number, and $a=1,2$, corresponding for example to in versus out, and we expect there to be symplectic such bases, with
\beq
\label{sympcond}
\int_\Sigma d^d y\, \phi^a_{klm} \pi^{a'}_{k'l'm'} = \delta_{aa'}\delta_{ll'} \delta_{mm'}\delta(k-k')\ \ .
\eeq
These modes can be defined in terms of functions of $r$,
\beq
\phi^a_{klm} = f^a_{kl}(r) Y_{lm}\quad ,\quad \pi^a_{klm} = g^a_{kl}(r) \sin\theta Y_{lm}\ ,
\eeq
and here we describe the case $D=4$ to simplify the angular notation, with the statements readily generalizing to higher $D$.
The functions $f^a_{kl}$, $g^a_{kl}$ should give a complete basis, satisfying the condition \eqref{sympcond}.

The problem is then that of describing a complex structure (choice of ``positive frequency") satisfying the local conditions for lying in the physical equivalence class described in Sec.~\ref{Localeq}, which correspondingly leads to a Hadamard two-point function.  As we have noted there are different such complex structures, and the condition to lie in the correct physical equivalence class is a short-distance condition on the modes.  If we consider the behavior near a point $r_0$, and assume that near that point the modes with large $k,l$ behave as
\beq
f^a_{kl}\sim e^{\pm ikr}\quad ,\quad g^a_{kl}\sim e^{\pm ikr}\ ,
\eeq
then the local positive-frequency behavior \eqref{CDH} corresponds to
\beq
\gamma^\pm_{klm}\propto e^{\pm ikr} Y_{lm}(1,-i\hat\omega_{kl} \sqrt{q_{rr}} r^2 \sin\theta)\ ,
\eeq
with
\beq
\hat\omega_{kl} =\sqrt{q^{rr}k^2 +\frac{l^2}{r^2}} \ ,
\eeq
in that limit.

Such modes are then expected to yield a Schr\"odinger picture of the standard form, as described in \cite{GiPe1}.  Specifically, the Cauchy data $\gamma^\pm_{klm}$ determines solutions $\varphi^\pm_{klm}(t,x)$ in the future of the slice $\Sigma$.  The local positive-frequency condition imposes the Hadamard condition on the two-point function in the neighborhood of $\Sigma$.  Then, if we assume the result of \cite{FNW}, preservation of the Hadamard condition under evolution, extends to spacetimes of stationary form, like those arising from the slicing \eqref{slicedef}, the Hadamard condition is satisfied in the future evolution.  If we consider a  subsequent timeslice $\Sigma'$, the corresponding solutions induce Cauchy data on this slice.  Since the corresponding two-point function is Hadamard, as is the two-point function arising from the Cauchy data $\gamma^\pm_{klm}$ on the slice $\Sigma'$, by stationarity, the corresponding complex structures should lie in the same equivalence class, that is, be related by a unitary transformation.  This transformation gives the time evolution of Schr\"odinger picture.  We of course see that this argument relies on  stationarity, and so does not necessarily go through in its absence.  Further aspects of such Schr\"odinger evolution were discussed in \cite{GiPe1}.

One can also study evolution on non-stationary geometries, which yields more complicated time evolution. An intermediary is to study a stationary black hole but in a non-stationary coordinate system.  In either case the conditions for modes to correspond to a complex structure in the physical equivalence class then becomes more subtle, and the modes have nontrivial time dependence. 
To study the near-horizon behavior, one can 
 use Kruskal coordinates,
\beq
X^{\pm}(t, r) = \pm 2R e^{\pm x^{\pm}(t, r)/2R} = T\pm X\ ,
\eeq
with $x^+ = x^-+2r_*$ and $r_*$ being the tortoise coordinate
\beq
r_*=r + R\ln\Big\vert\frac{r}{R} -1\Big\vert\ 
\eeq
(here given for $D=4$).
In the near-horizon ``Rindler region" where $|r-R| \ll R$, we can define Kruskal modes that behave approximately as modes in flat space with an effective mass parameter of $m = \sqrt{l(l+ 1)+ 1}/R$, which is detailed in section 6 and appendix B of \cite{GiPe1}. At any time $T$ 
the field and its momentum density is simply
\beq
\gamma^\pm_{klm}\propto e^{- i \omega_{kl} T \pm i k X} Y_{lm}(1,-i\omega_{kl} R^2 \sin\theta)\ ,
\eeq
where $\omega_{kl}^2 = k^2 +[l(l+ 1)+ 1]/R^2$, giving a good local definition of positive frequency modes.  However, extension of these modes outside this region becomes nontrivial.

\section{General discussion: implications, questions, and further directions}

We finish this paper with some discussion of the issues we have investigated, and their role in the problem of determining the evolving wavefunction, in particular when we consider interacting theories and dynamical geometry.

\subsection{Implications and questions: free theories}

While much of the initial discussion of issues with evolution on time-dependent slicings   focussed on resulting difficulties for passing to Schr\"odinger picture\cite{ToVa2,AgAs}, we first emphasize that the problem is a more general one involving the description of  the evolving quantum state.  We expect the basic issues to arise when 
we consider the general problem of a spacetime with initial and final slices $\Sigma_i$, $\Sigma_f$, not related through translation by a Killing vector.  In essence, we cannot choose a common basis defining the complex structure, {\it aka} ``choice of positive frequency modes," on both such embeddings of the general Cauchy slice $\Sigma$.  
If we consider evolution of free fields, as described above, 
in order for the evolved state at $\Sigma_f$ to be unitarily related to the initial state at $\Sigma_i$, it must be described in a class of complex structures that depend explicitly on  the embedding of $\Sigma$.  

Specifically, that means that the basic observables $\phi$, $\pi$ are related to the annihilation and creation operators of the Fock space construction of the Hilbert space through an expansion \eqref{ladderexp} that is not independent of the embedding.  This generalizes the corresponding explicit time dependence in the example of Robertson-Walker cosmology, described in \cite{AgAs},  also studied in \cite{MuOe2022}, and reviewed above.  Stated differently, a generic representation of the basic commutators \eqref{CCRs} will not lie in the physical equivalence class, and the evolution takes one out of any specified equivalence class.
So, when one considers a foliation interpolating between $\Sigma_i$ and $\Sigma_f$,  this  implies corresponding trouble for defining a standard Schr\"odinger picture evolution with a fixed representation of the basic operators $\phi$ and $\pi$.

This raises the question of defining the equivalence class of complex structures in which evolution between such slices {\it is} implemented by a unitary transformation.  
If we focus on free fields in a closed universe, to avoid infrared issues, Ref.~\cite{Waldbook} summarizes arguments for the existence of a unique equivalence class of such representations, determined by the condition that the two-point function be Hadamard.\footnote{As emphasized in \cite{Waldbook,Wald2009}, for noncompact Cauchy slices one does not have such a unique equivalence class.  These points are also nicely described in \cite{WittQFT}.}  The Hadamard condition can be specified in the vicinity of the initial slice $\Sigma_i$, and then is preserved under evolution\cite{FSW}, given a sufficiently continuous geometry.\footnote{One remaining question is to establish the relevant continuity condition.}

Particularly if one is to extend to interacting fields, or dynamical geometries  (see discussion below), it seems  desirable to have a condition for choosing the class of complex structures that does not depend on solving for the evolution.  We have investigated such conditions by examining the local short-distance behavior of evolution, and the corresponding positive-frequency condition, at short distance scales.  We have found conditions for the basis to have this local positive-frequency structure, and have also directly related that to the absence of non-Hadamard singularities.  Such a necessary condition on the local behavior of the complex structures was found to depend on the embedding of a Cauchy slice through the metric $q_{ij}$ induced on that slice, as described in Sec.~\ref{Localeq}.  This then specifies
a local (class of) Fock space constructions of the Hilbert space, related by unitary transformations satisfying the Hilbert-Schmidt condition.  These may have different particle number, but not {\it infinitely} different particle number.

One can connect this story to the existence of different ``generalized pictures" for evolution, with additional nontrivial behavior.  In a traditional Heisenberg picture description, the state is fixed and operators, such as $\phi$ and $\pi$, evolve with time.  The preceding arguments tell us that in curved backgrounds, unlike in standard quantum mechanics, there is not a unitary operator that transfers all of the time dependence to the state, and so that the operators $\phi$ and $\pi$ are time independent.  However, there are different unitarily-related complex structures, within the physical equivalence class, and so one can transfer different amounts of the time evolution from the operators $\phi$ and $\pi$ to the state, using the corresponding unitary transformations.  One can think of this as defining different generalized pictures, with the proviso that a standard Schr\"odinger picture with time independent operators does not in general exist.

Since, to rephrase the above, the problem with describing Schr\"odinger picture evolution lies in the unavoidable embedding, or time, dependence of the relation of the basic observables $\phi$ and $\pi$ to the annihilation/creation operators $a_A$, $a_A^\dagger$ used to define the states, that does suggest that one might treat the latter as fixed ``Schr\"odinger" operators, and define the hamiltonian and evolution of the state in terms of these.  But then, one has the additional embedding or time dependence arising in the relation of $\phi$ and $\pi$ to these operators, departing from a standard Schr\"odinger picture description.

Since the basic issues are ultraviolet issues, that raises the question of the possible role of a cutoff.  Of course, the problem of reconciling such a cutoff with even Lorentz invariance is a notorious one.  Possibly, however, if {\it e.g.} gravitational effects (see below) give a more subtle physical cutoff, that could evade the problem of inequivalent representations.

One approach to this set of problems has been the algebraic approach, see {\it e.g.} \cite{Haag}, where one focusses on the behavior of the operator algebras.  However, that is most clearly described in the context of free evolution, where one can solve the equations of motion to determine the operator evolution, and moreover leaves the question of what determines the state.  These limitations also suggest the importance of a local approach to investigating the state, such as we have described.

This discussion also raises the question of the role of such issues in a path integral formulation.  Suppose one wants to define transition amplitudes between slices $\Sigma_i$ and $\Sigma_f$ in terms of a functional integral over fields.  Quite plausibly, if the initial and final states are defined using the induced metrics $q_i$, $q_f$ to determine the representation, then the path integral gives the unitary evolution we have discussed in the operator approach; we will leave closer investigation of this to future work.

\subsection{Interacting theories with fixed backgrounds}

Most of the explicit discussion of these issues has been for free fields, and interacting theories present additional complications, which also appear to motivate a local approach like described in this paper.

For example, if one adopts the algebraic approach, then one expects the time evolution of the operators to be determined by the Heisenberg equations of motion.  However, when there are interactions these are no longer straightforward to solve, or even to precisely formulate.  That also leaves the question, noted above, of what determines the state.  

If one takes the viewpoint that the Hadamard condition should be used to determine the state and its evolution, one also encounters new challenges in the interacting case.  Specifically, in the preceding discussion one needed statements about evolution of the correlators, and specifically of the two-point function.  With interactions, this is likewise nontrivial, and makes the question of proving preservation of their Hadamard behavior also nontrivial.  

The challenge of solving the interacting equations of motion suggests the possible utility of instead having a Schr\"odinger picture description of evolution, which would  in principle provides an alternative to solving the Heisenberg equations.  But this does necessitate addressing the question of determining the physical representation of the commutators \eqref{CCRs} and accounting for its evolution.  The difficulty with doing so in the preceding approaches, which is also based on solving the equations of motion, suggests that a local approach to determining the representation of the quantum state could  be particularly useful and important in the interacting case; related comments were made in \cite{FrHa2}, making contact with the idea of local stability\cite{HNS}.

Specifically, at least in the case of an asymptotically free theory, we expect that the local conditions on the representation described in Sec.~\ref{Localsec} can also be used to determine the representation in the interacting theory.  This is expected to work because the condition for a physical representation is an asymptotic condition in the UV, and in this limit the interacting theory matches the free theory.  The very short time evolution is also expected to be that of the free theory, but over longer times the interacting evolution  builds up a nontrivial state.

\subsection{Dynamical geometry and the geometrical wavefunction}

To rephrase part of the preceding discussion, given a time slicing of a spacetime, one can relate the states on initial and final slices by a unitary transformation,
\beq
|\psi(t_f)\rangle = U(t_f,t_i)|\psi(t_i)\rangle\ .
\eeq
A simple example is in Heisenberg picture where $U=1$, but we may wish to describe the time evolution as that of the state, and eliminate it from the operators, as in Schr\"odinger picture.  Here we encounter an obstacle, since we find that it is in general not possible to describe the evolution with a fixed representation of the canonical commutators \eqref{CCRy}, with time {\it independent} $\pi$ and $\phi$.  Instead, the representation, aka relation between these operators and the annihilation/creation operators, depends on the slicing and the time within that slicing.  We do expect to be able to write the Heisenberg equations for evolving operators as
\beq
i\partial_t {\calo} = [H_m,\calo]\ ,
\eeq
with hamiltonian
\beq
H_m= \int d^3x\left[ \frac{N}{2}\left(\frac{\Pi^2}{\sqrt q} + \sqrt q q^{ij}\partial_i\varphi\partial_j\varphi\right) + N^i \Pi \partial_i\varphi\right] \ ,
\eeq
and use this to in particular determine the evolution of Heisenberg operators $\Pi(t)$ and $\varphi(t)$.  However, while this formally seems to define a unitary evolution operator
\beq
U(t,t_i) = e^{-i \int^t_{t_i} dt H_m}\ ,
\eeq
this can't be used to unitarily transfer the evolution to the state, {\it e.g.} by defining Schr\"odinger picture operators via
\beq
\calo_S = U(t,t_i)  \calo(t) U^\dagger(t,t_i) \ ,
\eeq
and correspondingly this appears to spell trouble for the Schr\"odinger evolution
\beq \label{Hevol}
i\partial_t |\psi\rangle = H |\psi\rangle
\eeq
with $H=H_m$ that one would na\"\i vely expect by comparison with flat-space field theory.  These do not account for the time dependence of the representation induced by the slicing.

These challenges appear to be magnified in quantum gravity, where one considers dynamical geometry.  If one considers evolution in a spacetime with an asymptotic region such as flat or AdS space, the full hamiltonian includes a gravitational contribution, $H=H_m+H_g$, and takes the form
\beq
H= \int d^3 x \left(N\calc_n + N^i\calc_i\right) + H_\partial\  .
\eeq
  Here $\calc_\mu$ are the gravitational constraints,
\beq
\calc_\mu = \sqrt q \left( -\frac{G_{\mu\nu}}{8\pi G} + T_{\mu\nu} \right) n^\nu\ ,
\eeq
and $H_\partial$ is an ADM boundary term, as  are for example reviewed in \cite{GiPe2}.

Na\"\i vely we expect evolution of the form \eqref{Hevol}, and given the relation of the hamiltonian to the constraints, we might expect this reduces to 
\beq
i\partial_t |\psi\rangle = H_\partial |\psi\rangle
\eeq
for states annihilated by the constraints.\footnote{Another subtlety with this is that only half of the constraints should annihilate the physical states\cite{DoGi1,GiPe2}.}  And, in a closed spacetime with no asymptotic region, we instead might expect the state to be determined by the condition that it is annihilated by the constraints, including the Wheeler-DeWitt equation.  But, in trying to solve these conditions, we seem to run into an analogous problem that the representation of the operators $\Pi$ and $\varphi$, and likewise now of the metric $q_{ij}$ and its conjugate momentum $P^{ij}$, which also enter the constraints 
\beq 
\calc_n = \hf\left(\frac{\Pi^2}{\sqrt q} + \sqrt q q^{ij}\partial_i\varphi\partial_j\varphi\right) - \frac{2\sqrt q}{\kappa^2} R_q + \frac{\kappa^2}{2\sqrt q}\left( P^{ij} P_{ij} -\frac{P^2}{D-2}\right)\ ,
\eeq
\beq
\calc_i=\Pi\partial_i\varphi -2 D_jP_i^j\ ,
\eeq
here with $D_i$ and $R_q$ the covariant derivative and curvature associated to $q_{ij}$, 
are not slice dependent, and instead introduce dependence on the slicing through the spatial metric $q$.  The fact that $q_{ij}$ is now an operator seems to amplify the difficulty.  Likewise, we expect obstacles to the na\"\i ve statement of the invariance under changes of many-fingered time,
\beq
|\psi\rangle \simeq e^{-i\int d^4x  (N \calc_n + N^i \calc_i)}|\psi\rangle
\eeq
arising from changes of slicing that differ in a compact region, but match asymptotically, and are parameterized by $N$ and $N^i$.  Related concerns were expressed in \cite{ToVa2}, perhaps too pessimistically.  
At the level of the algebra, we do expect the gravitational version of the Heisenberg equations to still hold, together with gauge invariance enforced through commutation with the $C_\mu$, determining for example gravitational dressings\cite{DoGi1},\cite{GiPe2}.  Of course, once one has dynamical quantum geometry, that also raises the question of a fundamental UV cutoff due to gravitational effects, {\it e.g.} as characterized by a ``locality bound\cite{GiLi2}," which could significantly modify the previous discussion of inequivalent representations.
We hope to return to these questions in future work.

\vskip.3in
\noindent{\bf Acknowledgements} 

This material is based upon work supported in part by the U.S. Department of Energy, Office of Science, under Award Number {DE-SC}0011702, by Heising-Simons Foundation grants \#2021-2819 and  \#2024-5307,  and by a Graduate Division Dissertation Fellowship at UCSB.  Part of this work was carried out at the Aspen Center for Physics, which is supported by National Science Foundation grant PHY-2210452.  SBG thanks G. Horowitz and D. Marolf for discussions, and J. Hartle for earlier discussions of subtleties of many-fingered time.

\appendix
\section{Finite volume quantization}
\label{appa}

This appendix will give conventions for quantization in finite volume, and extend those to treat the case where one considers quantization of fields restricted to a limited volume of spacetime, as for example in the local analysis of Sec.~\ref{Localsec}.

We begin with Minkowski space, with a periodic identification $x^i\simeq x^i +L$ in all $d$ spatial directions.  This implies quantized momenta, with
\beq
\label{kton}
k_i = \frac{2\pi n_i}{L}
\eeq
for integer $n_i$. Then, plane wave solutions
\beq
\label{Minksol}
\varphi_k = \frac{1}{\sqrt{2 \omega_k V}} e^{ik_\mu x^\mu} \ ,
\eeq
 with $k_0=-\omega_k=-\sqrt{k_i^2+m^2}$ and with volume $V=L^d$, will correspond to inner product
 \beq
 \label{innerp}
 \langle \gamma_k, \gamma^*_{k'}\rangle = \delta_{nn'} = \delta_{kk'}
 \eeq
 with $d$-dimensional Kronecker delta normalization.  Correspondingly, the field expansion
 \beq
\varphi =\sum_k\left(a_k \varphi_k + a_k^\dagger \varphi^*_k\right)
\eeq
gives
\beq
[a_k,a^\dagger_{k'}]= \delta_{kk'}\ .
\eeq
Note that we may take the infinite volume limit using
\beq
\label{infvol}
\sum_k \rightarrow \frac{V}{(2\pi)^d}\int d^dk\ ,
\eeq
from \eqref{kton}, and  correspondingly changing the scales of $\varphi_k$ and $a_k$.

We may introduce similar conventions to deal with a subregion with finite spatial volume, as in {\it e.g.} Sec.~\ref{Localsec}.  For example, if we have a partition of unity $\sum_I f_I(y)=1$ corresponding to different spatial regions with size $L'$, then the Cauchy data $\gamma_{Ik} \propto f_I(y) \gamma_k(y)$, arising from restricting the Cauchy data $\gamma_k$ of \eqref{Minksol}, is expected to give a useful basis for describing physics in region $I$ at distances $\ll L'$.  If we work on scales $L'$ small as compared to the curvature radius and that of the slices, region $I$ has volume $ V_I \simeq L^{'d} \sqrt {q(y_I)}$, with $y_I$ being a point near the center of the region. To maintain the unit normalization 
\eqref{innerp}, the modes restricted to region $I$ are  normalized analogously to \eqref{Minksol} as
\beq
\varphi_{Ik}\simeq \frac{1}{\sqrt{2 \hat \omega_k V_I}} e^{ik_a X^a}\ ,
\eeq
where now $X^a$ are the local flat coordinates, as defined in Sec.~\ref{Localeq}.

\mciteSetMidEndSepPunct{}{\ifmciteBstWouldAddEndPunct.\else\fi}{\relax}
\bibliographystyle{utphys}
\bibliography{timedep}{}

\end{document}